\documentclass{INTERSPEECH2023}
\usepackage{float}
\usepackage{multirow}
\usepackage{algorithm}
\usepackage{algpseudocode}
\usepackage{amsmath}
\usepackage{url}
\usepackage{algorithmicx}
\usepackage{soul}
\usepackage{colortbl}

\usepackage{xcolor}
\usepackage{adjustbox}

\usepackage{pifont}
\newcommand{\cmark}{\ding{51}}%
\newcommand{\xmark}{\ding{55}}%

\interspeechcameraready

\title{A Neural State-Space Model Approach to Efficient Speech Separation}
\name{Chen Chen$^1$, Chao-Han Huck Yang$^2$, Kai Li$^3$, Yuchen Hu$^1$, Pin-Jui	Ku$^3$, Eng Siong Chng$^1$}
\address{
  $^1$Nanyang Technological University, Singapore\\
  $^2$Georgia Institute of Technology, USA \\
  $^3$Tsinghua University, China}
\email{chen1436@e.ntu.edu.sg}

\begin{document}

\maketitle
 
\begin{abstract}
In this work, we introduce S4M, a new efficient speech separation framework based on neural state-space models (SSM). Motivated by linear time-invariant systems for sequence modeling, our SSM-based approach can efficiently model input signals into a format of linear ordinary differential equations (ODEs) for representation learning. To extend the SSM technique into speech separation tasks, we first decompose the input mixture into multi-scale representations with different resolutions. This mechanism enables S4M to learn globally coherent separation and reconstruction. The experimental results show that S4M performs comparably to other separation backbones in terms of SI-SDRi, while having a much lower model complexity with significantly fewer trainable parameters. In addition, our S4M-tiny model (1.8M parameters) even surpasses attention-based Sepformer (26.0M parameters) in noisy conditions with only 9.2\% of multiply-accumulate operation (MACs).
\end{abstract}

\section{Introduction}
Speech separation (SS) aims to separate target speech from overlapping speech signal sources~\cite{haykin2005cocktail}, also known as \emph{cocktail party problem}. SS widely serve as a pre-processor for speech applications~\cite{wang2018supervised, li2022use}, e.g., automatic speech recognition~\cite{yu2016automatic,hu2023gradient} and speaker verification~\cite{rosenberg1976automatic}. Recently, SS has gained remarkable progress driven by the power of deep learning~\cite{lecun2015deep,hu2023unifying,zhang2023noise}, where the clean speech of individual speakers serves as ground truth to supervise the training of the neural network~\cite{huang2014deep}.\par

Developing an efficient SS architecture with low model complexity is challenging due to the high-dimensional input of speech signals, which contains tens of thousands of time steps per second and exhibits long-range behaviors at multiple timescales. In order to handle this challenge, previous deep learning-based attempts have tailored standard sequence modeling approaches like CNNs~\cite{luo2019conv}, RNNs~\cite{luo2020dual,li2023design}, and Transformers~\cite{subakan2021attention} to predict clean speech from a mixture. However, these works have different limitations.CNNs are constrained by the size of the receptive field, making it difficult to achieve global coherence~\cite{goel2022s}. RNNs lack computational efficiency because they cannot be parallelized during training. While Transformers-based~\cite{vaswani2017attention} architectures achieve impressive performance on a public dataset, their vast network size (e.g., Sepformer with 26.0M parameters~\cite{subakan2021attention}) results in high computational costs for training and inference, hampering the application of the trained model in practical scenarios.  \par 

To improve the efficiency for SS, we are inspired by the recent advances in neural state-space model (SSM)~\cite{gu2021efficiently}, which have shown outstanding performance in high-rate audio generation tasks~\cite{goel2022s}. The globally coherent generation of SSM is similar to self-attention mechanism in Transformers, but with significantly fewer trainable parameters are required in SSM. Consequently, we believe that SSM offers a solution to reduce the model complexity of SS, thus improving the separation efficiency for both training and inference. \par 

In this paper, we introduce an efficient SS method called S4M (speech separation using state-space model), which follows the mainstream encoder-decoder pipeline. Specifically, the encoder in S4M extracts multiple features with varying resolutions from a flat input mixture, and then feeds them into S4 blocks to capture the representation with global long-range dependencies. Similarly, S4 layer is also employed in the decoder for feature reconstruction. The main strengths of S4M are summarized as follows:
\begin{itemize}
    \item S4M offers significant advantages over mainstream SS methods, in terms of model complexity and computational cost.  
    \item S4M effectively captures long-range dependencies for high-rate waveforms, which benefits separated feature reconstruction, especially in noisy conditions. 
\end{itemize}
To demonstrate these strengths of S4M, we conducted experiments on clean datasets WSJ0-2Mix and LibriMix, as well as the noisy dataset LRS2-Mix. The experimental results show that S4M achieves comparable performance with other competitive baselines in clean conditions, and achieves state-of-the-art performance on LRS2-Mix, which includes practical noise and reverberation in the mixture. Furthermore, we compared the model complexity of S4M with other models, and results show that S4M has remarkable superiority in terms of computational cost and inference time, making it one potential solution for streaming-based speech separations~\cite{wang2020voicefilter}. 
\begin{figure*}[t]
\begin{center}
\includegraphics[scale=0.83]{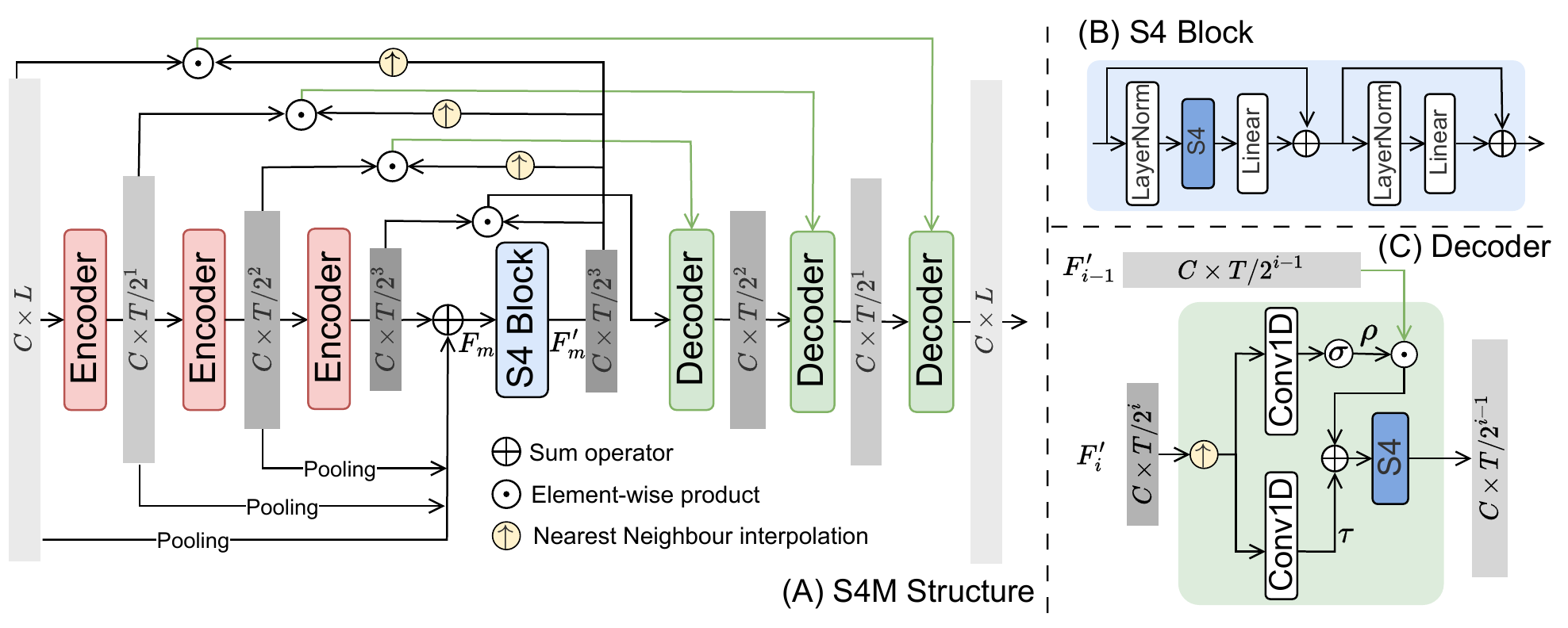}
\end{center}
\vspace{-0.1cm}
\caption{The block diagram of the (A) S4M model, (B) S4 Block, and (C) Decoder. ``$\sigma$" denotes the Sigmoid function. The grey chunks denote the hidden features after each layer.}
\vspace{-0.1cm}
\label{f1}
\end{figure*}

\section{Background: State-Space Models}
Given the input mixture $x\in \mathbb{R}^{1\times T}$, the goal of speech separation is to separate and predict clean speech $y^{n}\in \mathbb{R}^{1\times T}$ for each speaker, where $n$ is the number of speakers. CNNs and RNNs are the most widely used models for speech separation, each with its own advantages and limitations during training and inference. Specifically, a CNN layer computes a convolution with parameterized kernels
\begin{equation}
    K = (k_0, \cdots,k_{w-1}) \ \ \ \ \  \ \ \ y^{n} = K * x
\end{equation}
where $w$ is the width of the kernel. The receptive field or context size of a CNN is determined by the sum of kernel widths across all layers. As duration $T$ is usually large for speech signal, this results in increased computational complexity. To address this, a variant of CNNs called dilated convolution (DCNN) is widely used in SS, where each kernel K is non-zero only at its endpoints~\cite{yu2015multi}. On the other hand,  RNNs sequentially compute a hidden state $h_t$ from a previous history state $h_{t-1}$ and current input $x$. The output $y$ is modeled as:
\begin{equation}
    h_t = f(h_{t-1}, x) \ \ \ \ \  \ \ \ y = g(h_t)
\end{equation}
$f$ is also known as an RNN cell, such as the popular LSTM.
The recently proposed deep neural state-space model(SSM) advances speech tasks by combining the properties of both CNNs and RNNs. The SSM~\cite{gu2021efficiently} is defined in continuous time using the following equations:
\begin{align}
    & h'(t) = A h(t) + B x(t) \\
    & y(t) = C h(t) + D x(t)
\end{align}
To operate on discrete-time sequences sampled with a step size of $\Delta$, SSM can be computed with recurrence as follows:
\begin{align}
    & h_k = \overline{A}h_{k-1} + \overline{B}x_k \ \ \ \ \ y_k = \overline{C}h_k + \overline{D}x_k \label{eq5}\\
    & \overline{A} = (I-\Delta/2 \cdot A)^{-1} (I+\Delta/2 \cdot A) 
\end{align}
where $\overline{A},\overline{B}, \overline{C}, \overline{D}$ are the discretized state matrices. According to~\cite{goel2022s}, Eq.(\ref{eq5}) can be rewritten as a discrete convolution:
\begin{align}
     & y_k = \overline{CA}^k\overline{B}x_0 + \overline{CA}^{k-1}\overline{B}x_1 + \cdots + \overline{CB}x_k \\
     &y = \overline{K} * x  \ \ \ \ \ \ \  \overline{K} = (\overline{CB}, \overline{CAB}, \overline{C}A^2\overline{B}) \label{y=kx}   
\end{align}
$\overline{K}$ is the SSM convolution kernel.
The Eq.(\ref{y=kx}) is a single (non-circular) convolution and can be computed very efficiently with Fast Fourier Transformation, provided that $\overline{K}$ is known. \par
In order to calculate $\overline{K}$, we employ a specific instantiation of SSM, known as \textbf{S4 layer}~\cite{gu2021efficiently}, which parameterizes $A$ as a diagonal plus low-rank (DPLR) matrix: $A = \Lambda - pq^*$. This parameterization has three advantages: 1) Faster computation. The kernel $\overline{K}$ in Eq.(\ref{y=kx}) can be computed very quickly in this setting. 2) Improved capture of long-range dependencies. This parameterization includes HiPPO matrices~\cite{gu2020hippo}, which theoretically and empirically allow SSM to better capture global correspondence from input. 3) Better stability. SSM involves the spectrum of the state matrix $A$, which is more easily controlled since $-pp^*$ is always a negative semi-definite matrix~\cite{goel2022s}. \par
Given any time step $\Delta$, the computation of the SSM convolution kernel $\overline{K}$ requires $\mathcal{O}(S+L)$ operations and $\mathcal{O}(S+L)$ space, where $S$ is the state size and $L$ is the length of input.

\section{S4M: State-Space Speech Separation Model}
The overview structure of S4M is shown in Fig~\ref{f1}, where an encoder-decoder pipeline with S4 block is employed for speech separation tasks. As a time-domain method, S4M typically converts input waveform $x\in \mathbb{R}^{1\times T}$ to 2D features $F_0 \in \mathbb{R}^{C \times L}$ using a 1-D convolutional layer, where $C$ and $L$ represent the channel number and feature length respectively.  

\subsection{Encoder and S4 Block}
Prior works~\cite{chen2021time,li2022efficient} have demonstrated the advantages of using multi-scale representations with different resolutions for speech tasks. Consequently, we stack three down-sampling encoders (red blocks in Fig.~\ref{f1}) that consists of a 1-D dilation convolution layer, followed by a global normalization layer. The dilation factor is set as 2 to gradually increase the receptive field. In this way, the length dimension $L$ of feature is squeezed layer by layer, shown as the grey chunks in Fig.~\ref{f1}. Subsequently, a set of representations $F = \{F_i\in \mathbb{R}^{C \times \frac{L}{2^{i-1}}} |\ i=0, 1, \cdots  \} $ with same channel $C$ but different length $L$ are extracted from input, where $i$ is set as 4 in this paper. \par
To integrate the information from the multi-scale representations, we perform average pooling on the features from shallow layer to reshape them, and then add them to obtain feature $F_m \in \mathbb{R}^{C\times \frac{L}{8}}$. To capture global correspondence from $F_m$, a residual S4 block is employed (shown as blue box in Fig.\ref{f1}). Specifically, it contains a normalization layer, a S4 layer with GELU activation function~\cite{hendrycks2016gaussian}, and a linear layer. We also use additional point-wise linear layers in the style of a feed-forward network in Transformer, along with a residual connection to avoid the vanishing gradient problem. Notably, the S4 block does not change the shape of feature, therefore, the feature of $F'_m$ with shape of ${C\times \frac{L}{8}}$ is obtained after the S4 block.

\subsection{Decoder}
The decoder of S4M progressively reshapes the separated features to maintain symmetry with the encoder. As shown in Fig.\ref{f1}, two decoder inputs $F'_{i-1}$ and $F'_i$ are obtained by the element-wise multiplication between $F_{i-1}$ and $F'_m$, as well as $F_{i}$ and $F'_m$. This mask-based operation is commonly used in speech separation tasks. In addition, up-sampling of nearest neighbour interpolation is required for $F'_m$ due to shape mismatch. \par
Given $F_{i-1}$ and $F_i$, the decoder first employs a light local attention mechanism~\cite{li2022efficient} using adaptive parameters $\rho$ and $\tau$, which are respectively denoted as:
\begin{equation}
    \tau = f_2(\phi(F'_{i})) \ \ \  \ \ \ \rho = \sigma(f_1(\phi(F'_{i})))
\end{equation}
where $f_1$ and $f_2$ are two 1-D convolutional layers followed by normalization layer, $\phi$ denotes the nearest neighbor interpolation along time dimension $L$ for up-sampling ($C\times \frac{L}{2^{i}}$ $\rightarrow$ $C\times \frac{L}{2^{i-1}}$), and $\sigma$ denotes the Sigmoid function. As $F'_{i-1}$, $\rho$, and $\tau$ have the same shape, the local attention process is formulated by:
\begin{equation}
    F'_{i-1} = \rho \odot F'_{i-1} + \tau
\end{equation}
Then the same S4 block is employed for globally coherent generation after local attention. As shown in Fig.~\ref{f1}, the output of decoder is recursively multiplied by the output of encoder to get $F'_{i-2}$ and then fed it into next decoder layer until the output shape is restored to $C\times L$. \par
We adopt unfolding scheme for the network as proposed in A-FRCNN \cite{hu2021speech}. Concretely, the structure shown in Fig.~\ref{f1} (A) is repeated for B times (weight sharing), such that the input of the current model is also added by each previous model's output.
\subsection{Training objective}
The objective of training the end-to-end S4M is to maximize the scale-invariant source-to-noise ratio (SI-SNR), which is commonly used as the evaluation metric for source separation. The SI-SNR loss is defined as:
\begin{equation}
    \mathcal{L}_{si-snr} = -\sum_{n=1}^N10\log_{10}\left( {\frac{{{{\Vert {\frac{{{{\hat y}^T}_n{y_n}}}{{{{\Vert {{y_n}} \Vert}^2}}}{y_n}} \Vert}^2}}}{{{{\Vert {\frac{{{{\hat y}^T}_n{y_n}}}{{{{\Vert {{y_n}} \Vert}^2}}} - {{\hat y}_n}} \Vert}^2}}}} \right)
\end{equation}
where $y^{(n)}$ is the ground-truth signal for speaker $n$, and $\hat{y}^{(n)}$ is the estimated time-domain speech produced by S4M.  \par
Furthermore, Utterance-level permutation invariant training (uPIT) is applied during training to address the source permutation problem~\cite{kolbaek2017multitalker}. 

\begin{table}[t]
\centering
\resizebox{0.95\columnwidth}{!}{
\begin{tabular}{c|cc| c}
\toprule[1.5pt]
Model     & SI-SDRi (dB) & SDRi (dB) & \# Para. (M) \\ \midrule
ADANet~\cite{luo2018speaker}       & 9.1  & 10.4 & 9.1 \\
WA-MISI-5~\cite{wang2018end}    & 12.6 & 13.1 & 32.9 \\
SPN~\cite{wang2019deep}          & 15.3 & 15.6 & 56.6 \\
Conv-TasNet~\cite{luo2019conv}  & 15.3 & 15.6 &5.1 \\
Deep CASA~\cite{liu2019divide}    & 17.7 & 18.0 & 12.8 \\
FurcaNeXt~\cite{zhang2020furcanext}    & -    & 18.4 & 51.4 \\
TDANet ~\cite{li2022efficient}      & 18.6 & 18.9 & 2.3 \\
DPRNN~\cite{luo2020dual}        & 18.8 & 19.0 & 2.6 \\
SUDO RM-RF~\cite{tzinis2020sudo}   & 18.9 & - & 6.4 \\
Gated DPRNN~\cite{nachmani2020voice} & 20.1 & 20.4 & 7.5\\
Sepformer~\cite{subakan2021attention}    & 20.4 & 20.5 & 26.0 \\
Wavesplit~\cite{zeghidour2021wavesplit}    & 21.0 & 21.2 & 29.0 \\
SFSRNet~\cite{rixen2022sfsrnet}      & 22.0 & 22.1 & 59.0 \\
TF-GridNet~\cite{wang2022tf}   & \textbf{23.4} & \textbf{23.5} & 14.4 \\
\midrule
S4M-tiny    &  19.4 &  19.7 & \textbf{1.8}     \\
S4M         &  20.5 & 20.7 & 3.6       \\ 
\bottomrule[1.5pt]
\end{tabular}}
\vspace{0.2cm}
\caption{SI-SDRi and SDRi results on WSJ0-2Mix. ``\# Para." denotes the number of trainable parameters for each model. Best results are in bold.}
\vspace{-0.2cm}
\label{t0}
\end{table}

\begin{table}[]
\centering
\resizebox{0.96\columnwidth}{!}{
\begin{tabular}{c|cccc| c}
\toprule[1.5pt]
\multirow{2}{*}{Model} & \multicolumn{2}{c}{LibriMix} & \multicolumn{2}{c|}{LRS2-Mix} & \# Para.  \\ 
& SI-SDRi & SDRi & SI-SNRi & SDRi & (M) \\\midrule
BLSTM-TasNet      & 7.9 & 8.7 & 6.1 & 6.8 & 23.6  \\ 
Conv-TasNet       & 12.2 & 12.7 & 10.6 & 11.0 & 5.6  \\ 
DPRNN             & 16.1 & 16.6 & 12.7 & 13.0 & 2.7 \\
SuDoRM-RF            & 14.0 & 14.4 & 11.3 & 11.7 & 6.4 \\
Sepformer         & 16.5 & 17.0 & 13.5 & 13.8 & 26.0 \\
WaveSplit         & 16.6 & 17.2 & 13.1 &13.4 & 29.0 \\
A-FRCNN           & 16.7 & 17.2 & 13.0 & 13.3 & 6.1  \\ 
TDANet            & \textbf{17.4} & \textbf{17.9} & 14.2 & 14.5 & 2.3  \\ 
\midrule
S4M-tiny          & 16.2 &  16.6 &   14.2 & 14.5 & \textbf{1.8}        \\
S4M               & 16.9  & 17.4 &  \textbf{15.3} & \textbf{15.5}  & 3.6         \\ 
\bottomrule[1.5pt]
\end{tabular}}
\vspace{0.2cm}
\caption{SI-SDRi and SDRi results on LibriMix and LRS2-Mix.}
\vspace{-0.2cm}
\label{t1}
\end{table}
\section{Experiment}
\subsection{Database}
We evaluate S4M and other competitive methods on both clean and noisy datasets, including WSJ0-2Mix~\cite{hershey2016deep}, LibriMix~\cite{cosentino2020librimix} and LRS2-Mix~\cite{li2022efficient}. To ensure the generality, the mixture in test set are generated by the speakers that are not seen during training. \par
\noindent \textbf{WSJ0-2Mix} is the most common speech separation dataset derived from Wall Street Journal (WSJ0). It consists of a 30 hours of training set (20k utterances), a 8 hours of validation set (5k utterances), and a 5 hours of test set (3k utterances). All utterances are re-sampled to 8 kHz for comparison with other works.
\noindent \textbf{LibriMix}. Considering the limited data amount of WSJ0-2mix, we further employ LibriMix dataset to evaluate the performance in clean condition. The target speech in LibriMix is randomly drawn from the train-100 subset of LibriSpeech dataset with 8 kHz sampling rate. Each mixture uniformly samples Loudness Units relative to Full Scale (LUFS) between -25 and -33 dB. The training set contains 13.9k utterances with duration of 58 hours, while the validation set and test set both contain 3k utterances with duration of 11 hours. \par
\noindent \textbf{LRS2-Mix}. The source of LRS2-Mix is LRS2 dataset~\cite{chung2017lip} that includes thousands of video clips from BBC. It contains practical noise and reverberation interference, which is more close to reality. We randomly select utterances of 16 kHz from different scenes and mix them with signal-to-noise ratios sampled between -5 dB and 5dB. In practice, we utilize the same mixing script as WSJ0-2Mix, in which the training set, validation set and test set contain 20k, 5k and 3k utterances respectively.  \par
\subsection{S4M Setup}
The kernel size of convectional layer to process time domain signal is set as 4ms and the stride size is set as 1ms. The number of channels in the dilation convolution layer of the encoder and the number of hidden units in all linear layers are both set as 512. For S4 layer, we found that the model performs the best when the number of channels is set as 16. Furthermore, we develop a lighter version of S4M called \emph{S4M-tiny}, which removes the S4 layers (dark blue block in Fig.~\ref{f1}-C) in the decoder. It is worth noting that the S4M-tiny only contains 1.8M trainable parameters. \par
Both S4M and S4M-tiny are trained for 200 epochs with a learning rate of 0.001. An early-stopping strategy is adopted when validation loss does not decrease for 5 epochs. To avoid gradient explosion, we apply gradient clipping with a maximum L2 norm of 5 during training.

\subsection{Evaluation Metric}
We asses the clarity of separated audios based on scale-invariant signal-to-distortion ratio improvement (SI-SDRi) and signal-to-distortion ratio improvement (SDRi).
To evaluate model efficiency, we measure the processing time consumption per second for all models, indicated by real-time factor (RTF) in the tables. RTF is calculated by processing ten audio tracks of 1 second in length and 16 kHz in sample rate on CPU and GPU (total processing time / 10), represented as “CPU-RTF” and “GPU-RTF” respectively. The numbers are then averaged after running 1000 times. Also, we use the parameter size and the number of multiply-accumulate operations (MACs) to measure the model size. MACs are calculated using the open-source tool PyTorch-OpCounter4 under the MIT license.\par
For both SI-SDRi and SDRi, higher score indicates better quality of separated signal. For all efficiency metrics, lower value means lower complexity of model.

\section{Result and Analysis}
\subsection{Main results}
We report our main results on the test set of WSJ0-2Mix in Table~\ref{t0}, as well as LirbriMix and LRS3-Mix in Table~\ref{t1}. On WSJ0-2Mix dataset, S4M surpasses CNN-based Conv-TasNet and RNN-based DPRNN, and achieves comparable SI-SDRi performance with Transformer-based Sepformer, with far lower model complexity. In addition, S4M-tiny achieves 19.4 SI-SDRi performance with only 1.8M parameters, which demonstrates the efficiency of state-space model. For LibriMix dataset, we observe that S4M surpasses the Sepformer by 2.4\% on SI-SDRi performance when training data doubles (from 30 hours to 58 hours).   \par
We notice that S4M performs particularly well on LRS2-Mix which contains background noise and reverberation in the mixture. S4M-tiny even surpasses Sepformer by 13.3\% (13.5 dB $\rightarrow$ 15.3 dB) in terms of SI-SDRi, with only 6.9\% parameters. In addition, S4M achieves the best performance on LRS2-Mix in terms of both SI-SDRi and SDRi. This phenomenon indicates that S4M is effective to capture long-range dependencies for specific speaker, resulting in better noise-robustness in a more realistic environment.

\subsection{Ablation Study on S4}
We conduct ablation study on S4 module which serves as our main contribution. The results are summarized in Table~\ref{t2}. ``Mid." denotes whether S4 block exists between the encoder and decoder (Fig.~\ref{f1}-B), and  ``$S$" represents the dimension of the state in S4 block. ``Dec." denotes whether S4 layer is inserted in the decoder after local attention (Fig.~\ref{f1}-C), where the dimension of state is uniformly set as 16. \par
We observe that: 1) System 2 outperforms System 1 by a significant margin, highlighting the importance of S4 block which captures global correspondence from the multi-scale representations produced by the encoder and benefits subsequent separation. 2) The system achieves the best performance when the dimension of the state is set as 16. Increasing the value of ``$S$" leads to an increase in the number of parameters and a degradation in speech separation performance. 3) S4 layer is also effective for feature reconstruction in the decoder, but it inevitably increases the number of parameters.

\begin{table}[t]
\centering
\resizebox{0.95\columnwidth}{!}{
\begin{tabular}{c|cc|c|cc| c}
\toprule[1.5pt]
ID & Mid. & $S$ & Dec. & SI-SDRi & SDRi &\# Para.
\\ \midrule
1 & \xmark & - & \xmark  & 10.5 & 10.9 & 0.23 \\
2 & \cmark & 8 & \xmark  & 13.9 & 14.3  & 1.82 \\
3 & \cmark & 16 & \xmark & 14.2 & 14.5 & 1.84 \\
4 & \cmark & 32 & \xmark & 14.0 &  14.4  &  1.88 \\
5 & \cmark & 16 & \cmark & 15.3 &  15.5 & 3.59 \\
\bottomrule[1.5pt]
\end{tabular}}
\vspace{0.2cm}
\caption{Ablation study of S4 on LRS2-Mix dataset. ``\cmark" denotes S4 layer exist in corresponding module. ``\xmark" indicates the opposite. }
\label{t2}
\end{table}

\subsection{Analysis of model complexity}
We analyze the model complexity of S4M, which also indicates the separation efficiency. Using RTF and MACs as metrics, we report the performance of S4M and its comparison with other models in Table~\ref{t3}. GPU-RTF-\emph{f} and GPU-RTF-\emph{b} indicate the training time for each model on GPU devices, while CPU-RTF-\emph{f} denotes the inference speed on CPU devices when GPU resources is unavailable in some practical conditions. In addition, the ``TDANet (own)" is reproduced without accelerated Transformer by Pytorch.  \par

Table~\ref{t3} shows that S4M-tiny consistently requires the least training and inference time on both GPU and CPU devices. Moreover, S4M-tiny can achieve better performance than Sepformer on LRS2-Mix dataset, with only 9.2\% of MACs of Sepformer. For S4M, its model complexity is 4.8 times higher than S4M-tiny, but still significantly lower than Sepformer.  

\begin{table}[t]
\centering
\resizebox{0.98\columnwidth}{!}{
\begin{tabular}{c|cccc}
\toprule[1.5pt]
\multirow{2}{*}{Model} & GPU-RTF-\emph{f} & GPU-RTF-\emph{b}& CPU-RTF-\emph{f} & MACs \\
& (ms) & (ms) & (s) &(G/s)
\\ \midrule
BLSTM-TasNet & 233.85 & 654.14 & 5.90 & 43.0\\
SuDoRM-RF & 64.70 & 228.57 & 1.73 & 10.1 \\
DPRNN & 88.79 & 241.54 & 8.13 & 85.3 \\
A-FRCNN & 61.16 & 183.65 & 5.32 & 125.3 \\
TDANet & 23.77 & 97.92 & 1.78 & 9.1 \\ 
TDANet (own) & 61.25 & 368.54 & 5.97 & 9.1 \\ 
Sepformer & 65.61 & 184.91 & 7.55 & 86.9 \\ 
TF-GridNet & 100.52 & 285.37 & 86.4 & 128.7 \\ 
\midrule
S4M-tiny  & \textbf{18.19} & \textbf{73.62} & \textbf{1.34} & \textbf{8.0}                                 \\
S4M       & 40.15 & 132.83 & 2.57 & 38.7                                 \\
\bottomrule[1.5pt]
\end{tabular}}
\vspace{0.2cm}
\caption{Comparison of inference time and MACs on LRS2-Mix dataset.``-f" and ``-b" respectively stand for ``feed-forward" and ``backward" processes. For all metrics, lower is better.}
\vspace{-0.3cm}
\label{t3}
\end{table}

\section{Conclusion}
In this paper, we propose a efficient speech separation method (S4M) that achieves competitive performance while maintaining low model complexity. S4M utilizes state-space model to capture long-range dependencies from multi-scale representations, and integrates it into separated feature reconstruction. Experimental results show that S4M achieves comparable separation performance with significantly fewer trainable parameters in comparison with other mainstream methods. Furthermore, we analyze the model complexity using computing time and MACs, which shows that S4M provides a potential solution for streaming-based speech separation on mobile devices or streaming applications~\cite{hu2022transducer}.

\newpage

\bibliographystyle{IEEEtran}
\bibliography{mybib}

\end{document}